\documentclass[11pt] {article}

\usepackage{graphicx} 

\usepackage{amsmath,amssymb}

\title{A Three Function Variational Principle for Stationary Non-Barotropic Magnetohydrodynamics}

\author{Asher Yahalom$^{1,2}$ \\
$^1$Ariel University, Kiryat Hamada POB 3, Ariel 40700, Israel\\
$^2$Princeton University, Princeton, New Jersey 08543, USA\\
e-mails: asya@ariel.ac.il}

\begin{document}

\newcommand{\beq} {\begin{equation}}
\newcommand{\enq} {\end{equation}}
\newcommand{\ber} {\begin {eqnarray}}
\newcommand{\enr} {\end {eqnarray}}
\newcommand{\eq} {equation}
\newcommand{\eqn} {equation }
\newcommand{\eqs} {equations }
\newcommand{\ens} {equations}
\newcommand{\mn}  {{\mu \nu}}
\newcommand {\er}[1] {equation (\ref{#1}) }
\newcommand {\ern}[1] {equation (\ref{#1})}
\newcommand {\ers}[1] {equations (\ref{#1})}
\newcommand {\Er}[1] {Equation (\ref{#1}) }

\maketitle

\begin{abstract}
Variational principles for magnetohydrodynamics (MHD) were in\-troduced by
previous authors both in Lagrangian and Eulerian form. In this
paper we introduce simpler Eulerian variational principles from
which all the relevant equations of non-barotropic stationary magnetohydrodynamics can be derived for certain field topologies.
 The variational principle is given in terms of three independent functions
for stationary non-barotropic flows.
 This is a smaller number of variables than the eight
variables which appear in the standard equations of non-barotropic
magnetohydrodynamics which are the magnetic field $\vec B$ the
velocity field $\vec v$, the entropy $s$ and the density $\rho$. We further investigate the
case in the flow along magnetic lines is not ideal.
\end{abstract}

\section{Introduction}

Variational principles for magnetohydrodynamics were introduced by
previous authors both in Lagrangian and Eulerian form. Sturrock
\cite{Sturrock} has discussed in his book a Lagrangian variational
formalism for magnetohydrodynamics. Vladimirov and Moffatt
\cite{Moffatt} in a series of papers have discussed an Eulerian
variational principle for incompressible magnetohydrodynamics.
However, their variational principle contained three more
functions in addition to the seven variables which appear in the
standard equations of incompressible magnetohydrodynamics which are the magnetic
field $\vec B$ the velocity field $\vec v$ and the pressure $P$.
Kats \cite{Kats} has generalized Moffatt's work for compressible
non barotropic flows but without reducing the number of functions
and the computational load. Moreover, Kats has shown that the
variables he suggested can be utilized to describe the motion of
arbitrary discontinuity surfaces \cite{Kats3,Kats4}. Sakurai
\cite{Sakurai} has introduced a two function Eulerian variational
principle for force-free magnetohydrodynamics and used it as a
basis of a numerical scheme, his method is discussed in a book by
Sturrock \cite{Sturrock}. A method of solving the equations for
those two variables was introduced by Yang, Sturrock \& Antiochos
\cite{Yang}. Yahalom \& Lynden-Bell \cite{YaLy} combined the Lagrangian of
Sturrock \cite{Sturrock} with the Lagrangian of Sakurai
\cite{Sakurai} to obtain an {\bf Eulerian} Lagrangian principle for barotropic magnetohydrodynamics
which will depend on only six functions. The variational
derivative of this Lagrangian produced all the equations
needed to describe barotropic magnetohydrodynamics without any
additional constraints. The equations obtained resembled the
equations of Frenkel, Levich \& Stilman \cite{FLS} (see also \cite{Zakharov}).
Yahalom \cite{Yah} have shown that for the barotropic case four functions will
suffice. Moreover, it was shown that the cuts of some of those functions \cite{Yah2}
are topological local conserved quantities.

Previous work was concerned only with
barotropic magnetohydrodynamics. Variational principles of non
barotropic magnetohydrodynamics can be found in the work of
Bekenstein \& Oron \cite{Bekenstien} in terms of 15 functions and
V.A. Kats \cite{Kats} in terms of 20 functions. The author of
this paper suspect that this number can be somewhat reduced.
Moreover, A. V. Kats  in a remarkable paper \cite{Kats2} (section
IV,E) has shown that there is a large symmetry group (gauge
freedom)  associated with the choice of those functions, this
implies that the number of degrees of freedom can be reduced.
Yahalom \cite{Yahalomnb,Yahalomnb2} have shown that only five functions will suffice to
describe non barotropic magnetohydrodynamics in the case
that we enforce a Sakurai \cite{Sakurai} representation for the magnetic field.
Morrison \cite{Morrison} has suggested a Hamiltonian approach but this also depends on 8 canonical variables (see table 2 \cite{Morrison}).
The work of Yahalom \cite{Yahalomnb} was concerned with general non-stationary flows.
In a separate work \cite{Yahalomc} was concerned with stationary flows and introduced a 8 variable stationary variational principle, here we shall attempt to improve on this and obtain a 3 variable stationary variational principle for non-barotropic MHD. This will be done for a general
case in which the magnetic field lines need not lie on entropy surfaces, for the
 restricted case in which the magnetic field lines lie on entropy surfaces see \cite{Yahalomd}.

We anticipate applications of this study both to linear and
non-linear stability analysis of known non barotropic magnetohydrodynamic configurations \cite{VMI,AHH}
 and for designing efficient numerical schemes for integrating
the equations of fluid dynamics and magnetohydrodynamics
\cite{Yahalom,YahalomPinhasi,OphirYahPinhasKop}.
Another possible application is connected to  obtaining new analytic solutions in terms
of the variational variables \cite{Yah3}.

The plan of this paper is as follows: First we introduce the
standard notations and equations of non-barotropic
magnetohydrodynamics for the stationary and non-stationary cases.
Then we introduce the concepts of load and metage. The variational principle follows.

\section{Standard formulation of non-barotropic magnetohydrodynamics}

The standard set of \eqs solved for non-barotropic magnetohydrodynamics are given below:
\beq
\frac{\partial{\vec B}}{\partial t} = \vec \nabla \times (\vec v \times \vec B),
\label{Beq}
\enq
\beq
\vec \nabla \cdot \vec B =0,
\label{Bcon}
\enq
\beq
\frac{\partial{\rho}}{\partial t} + \vec \nabla \cdot (\rho \vec v ) = 0,
\label{masscon}
\enq
\beq
\rho \frac{d \vec v}{d t}=
\rho (\frac{\partial \vec v}{\partial t}+(\vec v \cdot \vec \nabla)\vec v)  = -\vec \nabla p (\rho,s) +
\frac{(\vec \nabla \times \vec B) \times \vec B}{4 \pi}.
\label{Euler}
\enq
\beq
 \frac{d s}{d t}=0.
\label{Ent}
\enq
The following notations are utilized: $\frac{\partial}{\partial t}$ is the temporal derivative,
$\frac{d}{d t}$ is the temporal material derivative and $\vec \nabla$ has its
standard meaning in vector calculus. $\vec B$ is the magnetic field vector, $\vec v$ is the
velocity field vector, $\rho$ is the fluid density and $s$ is the specific entropy. Finally $p (\rho,s)$ is the pressure which
depends on the density and entropy (the non-barotropic case).

The justification for those \eqs and the conditions under which they apply can be found in standard books on magnetohydrodynamics
(see for example \cite{Sturrock}).
The above applies to a collision-dominated plasma in local thermodynamic equilibrium.
Such conditions are seldom satisfied by physical plasmas, certainly not in astrophysics or in
 fusion-relevant magnetic confinement experiments. Never the less it is believed that the fastest
 macroscopic instabilities in those systems obey the above equations \cite{Yah2}, while instabilities
 associated with viscous or finite conductivity terms are slower. It should be noted that due to a
 theorem by Bateman \cite{Bateman} every physical system can be described by a variational principle (including viscous plasma) the
 trick is to find an elegant variational principle usually depending on a small amount of variational variables. The current
 work will discuss only ideal magnetohydrodynamics while viscous magnetohydrodynamics will be left for future endeavors.

 \Er{Beq}describes the
fact that the magnetic field lines are moving with the fluid elements ("frozen" magnetic field lines),
 \ern{Bcon} describes the fact that
the magnetic field is solenoidal, \ern{masscon} describes the conservation of mass and \ern{Euler}
is the Euler equation for a fluid in which both pressure
and Lorentz magnetic forces apply. The term:
\beq
\vec J =\frac{\vec \nabla \times \vec B}{4 \pi},
\label{J}
\enq
is the electric current density which is not connected to any mass flow.
\Er{Ent} describes the fact that heat is not created (zero viscosity, zero resistivity) in ideal non-barotropic magnetohydrodynamics
and is not conducted, thus only convection occurs.
The number of independent variables for which one needs to solve is eight
($\vec v,\vec B,\rho,s$) and the number of \eqs (\ref{Beq},\ref{masscon},\ref{Euler},\ref{Ent}) is also eight.
Notice that \ern{Bcon} is a condition on the initial $\vec B$ field and is satisfied automatically for
any other time due to \ern{Beq}. For the stationary case in which the physical fields do not depend on time
we obtain the following set of stationary equations:
\beq
\vec \nabla \times (\vec v \times \vec B)=0,
\label{Beqs}
\enq
\beq
\vec \nabla \cdot \vec B =0,
\label{Bcons}
\enq
\beq
 \vec \nabla \cdot (\rho \vec v ) = 0,
\label{masscons}
\enq
\beq
\rho (\vec v \cdot \vec \nabla)\vec v  = -\vec \nabla p (\rho,s) +
\frac{(\vec \nabla \times \vec B) \times \vec B}{4 \pi}.
\label{Eulers}
\enq
\beq
 \vec v \cdot \vec \nabla s = 0.
\label{Ents}
\enq

\section{Variational principle of non-barotropic magnetohydrodynamics}

In the following section we will generalize the approach of \cite{YaLy} for the non-barotropic case.
Consider the action:
\ber A & \equiv & \int {\cal L} d^3 x dt,
\nonumber \\
{\cal L} & \equiv & {\cal L}_1 + {\cal L}_2,
\nonumber \\
{\cal L}_1 & \equiv & \rho (\frac{1}{2} \vec v^2 - \varepsilon (\rho,s)) +  \frac{\vec B^2}{8 \pi},
\nonumber \\
{\cal L}_2 & \equiv & \nu [\frac{\partial{\rho}}{\partial t} + \vec \nabla \cdot (\rho \vec v )]
- \rho \alpha \frac{d \chi}{dt} - \rho \beta \frac{d \eta}{dt} - \rho \sigma \frac{d s}{dt}
 \nonumber \\
 &-& \frac{\vec B}{4 \pi} \cdot \vec \nabla \chi \times \vec \nabla \eta.
\label{Lagactionsimp}
\enr
In the above $\varepsilon$ is the specific internal energy (internal energy per unit of mass).
The reader is reminded of the following thermodynamic relations which
will become useful later:
\ber
d \varepsilon &=& T ds - P d \frac{1}{\rho} = T ds + \frac{P}{\rho^2} d \rho
\nonumber \\
& & \frac{\partial \varepsilon}{\partial s} = T, \qquad \frac{\partial \varepsilon}{\partial \rho} = \frac{P}{\rho^2}
\nonumber \\
w &=& \varepsilon + \frac{P}{\rho}= \varepsilon + \frac{\partial \varepsilon}{\partial \rho} \rho = \frac{\partial (\rho \varepsilon)}{\partial \rho}
\nonumber \\
dw &=& d\varepsilon + d(\frac{P}{\rho}) = T ds +  \frac{1}{\rho} dP
\label{thermodyn}
\enr
in the above $T$ is the temperature and $w$ is the specific enthalpy.
in the above: $\varepsilon$ is the specific internal energy, $T$ is the temperature and $w$ is the specific enthalpy. A special case of equation of state is the polytropic equation of state \cite{bt}:
\beq
p =K \rho^{\gamma}
\label{plytrop}
\enq
$K$ and $\gamma$ may depend on the specific entropy $s$. Hence:
\beq
\frac{\partial \varepsilon}{\partial \rho} = K \rho^{\gamma-2} \Rightarrow \varepsilon = \frac{K}{\gamma-1} \rho^{\gamma-1}
 =  \frac{p}{\rho(\gamma-1)}  \Rightarrow \rho \varepsilon = \frac{p}{\gamma-1}
\label{plytrop2}
\enq
the last identity is up to a function dependent on $s$.
Obviously $\nu,\alpha,\beta,\sigma$ are Lagrange multipliers which were inserted in such a
way that the variational principle will yield the following \ens:
\ber
& & \frac{\partial{\rho}}{\partial t} + \vec \nabla \cdot (\rho \vec v ) = 0,
\nonumber \\
& & \rho \frac{d \chi}{dt} = 0,
\nonumber \\
& & \rho \frac{d \eta}{dt} = 0.
\nonumber \\
& & \rho \frac{d s}{dt} = 0.
\label{lagmul}
\enr
It {\bf is not} assumed that $\nu,\alpha,\beta,\sigma$  are single valued.
Provided $\rho$ is not null those are just the continuity \ern{masscon}, entropy conservation
 and the conditions that Sakurai's functions are comoving.
Taking the variational derivative with respect to $\vec B$ we see that
\beq
\vec B = \hat {\vec B} \equiv \vec \nabla \chi \times \vec \nabla \eta.
\label{Bsakurai2}
\enq
Hence $\vec B$ is in Sakurai's form and satisfies \ern{Bcon}.
It can be easily shown that provided that $\vec B$ is in the form given in \ern{Bsakurai2},
and \ers{lagmul} are satisfied, then also \ern{Beq} is satisfied.

For the time being we have showed that all the equations of non-barotro\-pic magnetohydrodynamics can be obtained from the above variational principle except Euler's equations. We will now
show that Euler's equations can be derived from the above variational principle
as well. Let us take an arbitrary variational derivative of the above action with
respect to $\vec v$, this will result in:
\ber
\delta_{\vec v} A \hspace{-0.2cm} &=& \hspace{-0.2cm} \int dt \{ \int d^3 x dt \rho \delta \vec v \cdot
[\vec v - \vec \nabla \nu - \alpha \vec \nabla \chi - \beta \vec \nabla \eta - \sigma \vec \nabla s]
\nonumber \\
&+& \oint d \vec S \cdot \delta \vec v \rho \nu+  \int d \vec \Sigma \cdot \delta \vec v \rho [\nu]\}.
\label{delActionv}
\enr
The integral $\oint d \vec S \cdot \delta \vec v \rho \nu$ vanishes in many physical scenarios.
In the case of astrophysical flows this integral will vanish since $\rho=0$ on the flow
boundary, in the case of a fluid contained
in a vessel no flux boundary conditions $\delta \vec v \cdot \hat n =0$ are induced
($\hat n$ is a unit vector normal to the boundary). The surface integral $\int d \vec \Sigma$
 on the cut of $\nu$ vanishes in the case that $\nu$ is single valued and $[\nu]=0$ as is the case for
 some flow topologies. In the case that $\nu$ is not single valued only a Kutta type velocity perturbation \cite{YahPinhasKop} in which
the velocity perturbation is parallel to the cut will cause the cut integral to vanish. An arbitrary
velocity perturbation on the cut will indicate that $\rho=0$ on this surface which is contradictory to
the fact that a cut surface is to some degree arbitrary as is the case for the zero line of an azimuthal
angle. We will show later that the "cut" surface is co-moving with the flow hence it may become quite complicated.
This uneasy situation may be somewhat be less restrictive when the flow has some symmetry properties.

Provided that the surface integrals do vanish and that $\delta_{\vec v} A =0$ for an arbitrary
velocity perturbation we see that $\vec v$ must have the following form:
\beq
\vec v = \hat {\vec v} \equiv \vec \nabla \nu + \alpha \vec \nabla \chi + \beta \vec \nabla \eta + \sigma \vec \nabla s.
\label{vform}
\enq
The above equation is reminiscent of Clebsch representation in non magnetic fluids \cite{Clebsch1,Clebsch2}.
Let us now take the variational derivative with respect to the density $\rho$ we obtain:
\ber
\delta_{\rho} A & = & \int d^3 x dt \delta \rho
[\frac{1}{2} \vec v^2 - w  - \frac{\partial{\nu}}{\partial t} -  \vec v \cdot \vec \nabla \nu]
\nonumber \\
 & + & \int dt \oint d \vec S \cdot \vec v \delta \rho  \nu +
  \int dt \int d \vec \Sigma \cdot \vec v \delta \rho  [\nu]
  \nonumber \\
 & + & \int d^3 x \nu \delta \rho |^{t_1}_{t_0}.
\label{delActionrho}
\enr
In which $ w= \frac{\partial (\varepsilon \rho)}{\partial \rho}$ is the specific enthalpy.
Hence provided that $\oint d \vec S \cdot \vec v \delta \rho  \nu$ vanishes on the boundary of the domain
and $ \int d \vec \Sigma \cdot \vec v \delta \rho  [\nu]$ vanishes on the cut of $\nu$
in the case that $\nu$ is not single valued\footnote{Which entails either a Kutta type
condition for the velocity in contradiction to the "cut" being an arbitrary surface, or a vanishing density perturbation on the cut.}
and in initial and final times the following \eqn must be satisfied:
\beq
\frac{d \nu}{d t} = \frac{1}{2} \vec v^2 - w, \qquad
\label{nueq}
\enq
Since the right hand side of the above equation is single valued as it is made of physical quantities, we conclude that:
\beq
\frac{d [\nu]}{d t} = 0.
\label{mplicated nueqc}
\enq
Hence the cut value is co-moving with the flow and thus the cut surface may become arbitrary complicated.
This uneasy situation may be somewhat be less restrictive when the flow has some symmetry properties.

Finally we have to calculate the variation with respect to both $\chi$ and $\eta$
this will lead us to the following results:
\ber
\delta_{\chi} A \hspace{-0.4cm} & = & \hspace{-0.4cm} \int d^3 x dt \delta \chi
[\frac{\partial{(\rho \alpha)}}{\partial t} +  \vec \nabla \cdot (\rho \alpha \vec v)-
\vec \nabla \eta \cdot \vec J]
\nonumber \\
&+& \int dt \oint d \vec S \cdot [\frac{\vec B}{4 \pi} \times \vec \nabla \eta - \vec v \rho \alpha]\delta \chi
 \nonumber \\
 & + & \int dt \int d \vec \Sigma \cdot [\frac{\vec B}{4 \pi} \times \vec \nabla \eta - \vec v \rho \alpha][\delta \chi]
 \nonumber \\
 &-& \int d^3 x \rho \alpha \delta \chi |^{t_1}_{t_0},
\label{delActionchi}
\enr
\ber
\delta_{\eta} A \hspace{-0.4cm} & = & \hspace{-0.4cm} \int d^3 x dt \delta \eta
[\frac{\partial{(\rho \beta)}}{\partial t} +  \vec \nabla \cdot (\rho \beta \vec v)+
\vec \nabla \chi \cdot \vec J]
\nonumber \\
&+& \int dt \oint d \vec S \cdot [\vec \nabla \chi \times \frac{\vec B}{4 \pi} - \vec v \rho \beta]\delta \eta
\nonumber \\
 & + &  \int dt \int d \vec \Sigma \cdot [\vec \nabla \chi \times \frac{\vec B}{4 \pi} - \vec v \rho \beta][\delta \eta]
 \nonumber \\
 &-& \int d^3 x \rho \beta \delta \eta |^{t_1}_{t_0}.
\label{delActioneta}
\enr
Provided that the correct temporal and boundary conditions are met with
respect to the variations $\delta \chi$ and $\delta \eta$ on the domain boundary and
on the cuts in the case that some (or all) of the relevant functions are non single valued.
we obtain the following set of equations:
\beq
\frac{d \alpha}{dt} = \frac{\vec \nabla \eta \cdot \vec J}{\rho}, \qquad
\frac{d \beta}{dt} = -\frac{\vec \nabla \chi \cdot \vec J}{\rho},
\label{albetaeq}
\enq
in which the continuity \ern{masscon} was taken into account. By correct temporal conditions we
mean that both $\delta \eta$ and $\delta \chi$ vanish at initial and final times. As for boundary
conditions which are sufficient to make the boundary term vanish on can consider the case that
the boundary is at infinity and both $\vec B$ and $\rho$ vanish. Another possibility is that the boundary is
impermeable and perfectly conducting. A sufficient condition for the integral over the "cuts" to vanish
is to use variations $\delta \eta$ and $\delta \chi$ which are single valued. It can be shown that
$\chi$ can always be taken to be single valued, hence taking $\delta \chi$ to be single valued is no
restriction at all. In some topologies $\eta$ is not single valued and in those cases a single valued
restriction on $\delta \eta$ is sufficient to make the cut term null.

Finally we take a variational derivative with respect to the entropy $s$:
\ber
\delta_{s} A \hspace{-0.4cm} & = & \hspace{-0.4cm} \int d^3 x dt \delta s
[\frac{\partial{(\rho \sigma)}}{\partial t} +  \vec \nabla \cdot (\rho \sigma \vec v)- \rho T]
 \nonumber \\
&+& \int dt \oint d \vec S \cdot \rho \sigma \vec v  \delta s
-  \int d^3 x \rho \sigma \delta s |^{t_1}_{t_0},
\label{delActions}
\enr
in which the temperature is $T=\frac{\partial \varepsilon}{\partial s}$. We notice that according
to \ern{vform} $\sigma$ is single valued and hence no cuts are needed. Taking into account the continuity
\ern{masscon} we obtain for locations in which the density $\rho$ is not null the result:
\beq
\frac{d \sigma}{dt} =T,
\label{sigmaeq}
\enq
provided that $\delta_{s} A$ vanished for an arbitrary $\delta s$.

\section{Euler's equations}

We shall now show that a velocity field given by \ern{vform}, such that the
\eqs for $\alpha, \beta, \chi, \eta, \nu, \sigma, s$ satisfy the corresponding equations
(\ref{lagmul},\ref{nueq},\ref{albetaeq},\-\ref{sigmaeq}) must satisfy Euler's equations.
Let us calculate the material derivative of $\vec v$:
\ber
\frac{d\vec v}{dt} &=& \frac{d\vec \nabla \nu}{dt}  + \frac{d\alpha}{dt} \vec \nabla \chi +
 \alpha \frac{d\vec \nabla \chi}{dt}  +
\frac{d\beta}{dt} \vec \nabla \eta + \beta \frac{d\vec \nabla \eta}{dt}
\nonumber \\
&+&\frac{d\sigma}{dt} \vec \nabla s + \sigma \frac{d\vec \nabla s}{dt}.
\label{dvform}
\enr
It can be easily shown that:
\ber
\frac{d\vec \nabla \nu}{dt} & = & \vec \nabla \frac{d \nu}{dt}- \vec \nabla v_k \frac{\partial \nu}{\partial x_k}
 = \vec \nabla (\frac{1}{2} \vec v^2 - w)- \vec \nabla v_k \frac{\partial \nu}{\partial x_k},
 \nonumber \\
 \frac{d\vec \nabla \eta}{dt} & = & \vec \nabla \frac{d \eta}{dt}- \vec \nabla v_k \frac{\partial \eta}{\partial x_k}
 = - \vec \nabla v_k \frac{\partial \eta}{\partial x_k},
 \nonumber \\
 \frac{d\vec \nabla \chi}{dt} & = & \vec \nabla \frac{d \chi}{dt}- \vec \nabla v_k \frac{\partial \chi}{\partial x_k}
 = - \vec \nabla v_k \frac{\partial \chi}{\partial x_k},
  \nonumber \\
 \frac{d\vec \nabla s}{dt} & = & \vec \nabla \frac{d s}{dt}- \vec \nabla v_k \frac{\partial s}{\partial x_k}
 = - \vec \nabla v_k \frac{\partial s}{\partial x_k}.
 \label{dnabla}
\enr
In which $x_k$ is a Cartesian coordinate and a summation convention is assumed. Inserting the result from equations (\ref{dnabla},\ref{lagmul})
into \ern{dvform} yields:
\ber
\frac{d\vec v}{dt} &=& - \vec \nabla v_k (\frac{\partial \nu}{\partial x_k} + \alpha \frac{\partial \chi}{\partial x_k} +
\beta \frac{\partial \eta}{\partial x_k} + \sigma \frac{\partial s}{\partial x_k})
 \nonumber \\
&+& \vec \nabla (\frac{1}{2} \vec v^2 - w)+ T \vec \nabla s
 \nonumber \\
&+& \frac{1}{\rho} ((\vec \nabla \eta \cdot \vec J)\vec \nabla \chi - (\vec \nabla \chi \cdot \vec J)\vec \nabla \eta)
 \nonumber \\
&=& - \vec \nabla v_k v_k + \vec \nabla (\frac{1}{2} \vec v^2 - w) + T \vec \nabla s
\nonumber \\
&+& \frac{1}{\rho} \vec J \times (\vec \nabla \chi \times  \vec \nabla \eta)
 \nonumber \\
&=& - \frac{\vec \nabla p}{\rho} + \frac{1}{\rho} \vec J \times \vec B.
\label{dvform2}
\enr
In which we have used both \ern{vform} and \ern{Bsakurai2} in the above derivation. This of course
proves that the non-barotropic Euler equations can be derived from the action given in \er{Lagactionsimp} and hence
all the equations of non-barotropic magnetohydrodynamics can be derived from the above action
without restricting the variations in any way except on the relevant boundaries and cuts.

\section{Simplified action}

The reader of this paper might argue here that the paper is misleading. The author has declared
that he is going to present a simplified action for non-barotropic magnetohydrodynamics instead he
 added six more functions $\alpha,\beta,\chi,\-\eta,\nu,\sigma$ to the standard set $\vec B,\vec v,\rho,s$.
In the following I will show that this is not so and the action given in \ern{Lagactionsimp} in
a form suitable for a pedagogic presentation can indeed be simplified. It is easy to show
that the Lagrangian density appearing in \ern{Lagactionsimp} can be written in the form:
\ber
{\cal L} & = & -\rho [\frac{\partial{\nu}}{\partial t} + \alpha \frac{\partial{\chi}}{\partial t}
+ \beta \frac{\partial{\eta}}{\partial t}+ \sigma \frac{\partial{s}}{\partial t}+\varepsilon (\rho,s)]
\nonumber \\
&+& \frac{1}{2}\rho [(\vec v-\hat{\vec v})^2-(\hat{\vec v})^2]
\nonumber \\
& + &   \frac{1}{8 \pi} [(\vec B-\hat{\vec B})^2-(\hat{\vec B})^2]+
\frac{\partial{(\nu \rho)}}{\partial t} + \vec \nabla \cdot (\nu \rho \vec v ).
\label{Lagactionsimp4}
\enr
In which $\hat{\vec v}$ is a shorthand notation for $\vec \nabla \nu + \alpha \vec \nabla \chi +
 \beta \vec \nabla \eta +  \sigma \vec \nabla s $ (see \ern{vform}) and $\hat{\vec B}$ is a shorthand notation for
 $\vec \nabla \chi \times \vec \nabla \eta$ (see \ern{Bsakurai2}). Thus ${\cal L}$ has four contributions:
\ber
  {\cal L}  &  = &  \hat {\cal L} + {\cal L}_{\vec v}+ {\cal L}_{\vec B}+{\cal L}_{boundary},
\nonumber \\
\hat {\cal L}   & \equiv &   -\rho \left[\frac{\partial{\nu}}{\partial t} + \alpha \frac{\partial{\chi}}{\partial t}
+ \beta \frac{\partial{\eta}}{\partial t}+ \sigma \frac{\partial{s}}{\partial t}\right.
\nonumber \\
&+& \left. \varepsilon (\rho,s)+ \frac{1}{2} (\vec \nabla \nu + \alpha \vec \nabla \chi +  \beta \vec \nabla \eta +  \sigma \vec \nabla s )^2 \right]
\nonumber \\
&-&\frac{1}{8 \pi}(\vec \nabla \chi \times \vec \nabla \eta)^2
\nonumber \\
{\cal L}_{\vec v} &\equiv & \frac{1}{2}\rho (\vec v-\hat{\vec v})^2,
\nonumber \\
{\cal L}_{\vec B} &\equiv & \frac{1}{8 \pi} (\vec B-\hat{\vec B})^2,
\nonumber \\
{\cal L}_{boundary} &\equiv & \frac{\partial{(\nu \rho)}}{\partial t} + \vec \nabla \cdot (\nu \rho \vec v ).
\label{Lagactionsimp5}
\enr
The only term containing $\vec v$ is\footnote{${\cal L}_{boundary}$ also depends on
$\vec v$ but being a boundary term is space and time it does not contribute to the derived equations}
 ${\cal L}_{\vec v}$, it can easily be seen that
this term will lead, after we nullify the variational derivative with respect to $\vec v$,
to \ern{vform} but will otherwise
have no contribution to other variational derivatives. Similarly the only term containing $\vec B$
is ${\cal L}_{\vec B}$ and it can easily be seen that
this term will lead, after we nullify the variational derivative, to \ern{Bsakurai2} but will
have no contribution to other variational derivatives. Also notice that the term ${\cal L}_{boundary}$
contains only complete partial derivatives and thus can not contribute to the equations although
it can change the boundary conditions. Hence we see that \ers{lagmul}, \ern{nueq}, \ers{albetaeq} and \er{sigmaeq}
can be derived using the Lagrangian density:
\ber
& & \hspace{-1cm} \hat {\cal L}[\alpha,\beta,\chi,\eta,\nu,\rho,\sigma,s] = -\rho [\frac{\partial{\nu}}{\partial t} + \alpha \frac{\partial{\chi}}{\partial t}
+ \beta \frac{\partial{\eta}}{\partial t}+ \sigma \frac{\partial{s}}{\partial t}
\nonumber \\
&+ & \varepsilon (\rho,s) + \frac{1}{2} (\vec \nabla \nu + \alpha \vec \nabla \chi +  \beta \vec \nabla \eta +  \sigma \vec \nabla s )^2 ]
\nonumber \\
&-&\frac{1}{8 \pi}(\vec \nabla \chi \times \vec \nabla \eta)^2
\label{Lagactionsimp6}
\enr
in which $\hat{\vec v}$ replaces $\vec v$ and $\hat{\vec B}$ replaces $\vec B$ in the relevant equations.
Furthermore, after integrating the eight \eqs
(\ref{lagmul},\ref{nueq},\ref{albetaeq},\ref{sigmaeq}) we can insert the potentials $\alpha,\beta,\chi,\eta,\nu,\sigma,s$
into \ers{vform} and (\ref{Bsakurai2}) to obtain the physical quantities $\vec v$ and $\vec B$.
Hence, the general non-barotropic magnetohydrodynamic problem is reduced from eight equations
(\ref{Beq},\ref{masscon},\ref{Euler},\ref{Ent}) and the additional constraint (\ref{Bcon})
to a problem of eight first order (in the temporal derivative) unconstrained equations.
Moreover, the entire set of equations can be derived from the Lagrangian density $\hat {\cal L}$.

\section{Stationary non-barotropic MHD}
\label{Statmag}

Stationary flows are a unique phenomena of Eulerian fluid dynamics which has
no counter part in Lagrangian fluid dynamics. The stationary flow is defined
by the fact that the physical fields $\vec v,\vec B,\rho,s$ do not depend on the
temporal coordinate. This, however, does not imply that the corresponding potentials
$\alpha,\beta,\chi,\eta,\nu,\sigma$ are all functions of spatial coordinates alone.
Moreover, it can be shown that choosing the potentials in such a way will lead
to erroneous results in the sense that the stationary equations of motion can
not be derived from the Lagrangian density $\hat {\cal L}$ given in \ern{Lagactionsimp5}.
However, this problem can be amended easily
as follows. Let us choose $\alpha,\beta,\chi,\nu,\sigma$ to depend on the spatial coordinates alone.
Let us choose $\eta$ such that:
\beq
\eta = \bar \eta - t,
\label{etastas}
\enq
in which $\bar \eta$ is a function of the spatial coordinates. The Lagrangian
density $\hat {\cal L}$ given in \ern{Lagactionsimp5} will take the form:
\ber
\hat {\cal L} &=& \rho (\beta -\varepsilon (\rho,s))-
\frac{1}{2}\rho (\vec \nabla \nu + \alpha \vec \nabla \chi + \beta \vec \nabla \bar \eta+\sigma \vec \nabla s)^2
\nonumber \\
&-&\frac{1}{8 \pi}(\vec \nabla \chi \times \vec \nabla \bar \eta)^2.
\label{stathatL}
\enr
The above functional can be compared
with Vladimirov and Moffatt \cite{Moffatt} equation 6.12 for incompressible flows in which their
$I$ is analogue to our $\beta$. Notice however, that while $\beta$ is not a conserved quantity $I$ is.

Varying the Lagrangian $\hat {L} = \int \hat {\cal L} d^3x$ with respect to $\nu,\alpha,\beta,\chi,\eta,\rho,\sigma,s$
leads to the following equations:
\ber
& &  \vec \nabla \cdot (\rho \hat {\vec v} ) = 0,
\nonumber \\
& & \rho \hat {\vec v} \cdot \vec \nabla \chi = 0,
\nonumber \\
& & \rho (\hat {\vec v} \cdot \vec \nabla \bar \eta - 1) = 0,
\nonumber \\
& & \hat {\vec v} \cdot \vec \nabla \alpha = \frac{\vec \nabla \bar \eta \cdot \hat{\vec J}}{\rho}, \qquad
\nonumber \\
& & \hat {\vec v} \cdot \vec \nabla \beta = -\frac{\vec \nabla \chi \cdot \hat{\vec J}}{\rho},
\nonumber \\
& & \beta= \frac{1}{2} \hat {\vec v}^2 + w,
\nonumber \\
& & \rho \hat {\vec v} \cdot \vec \nabla s = 0,
\nonumber \\
& & \rho \hat {\vec v} \cdot \vec \nabla \sigma  =\rho T.
\label{statlagmul}
\enr
Calculations similar to the ones done in previous subsections will show that those equations
lead to the stationary non-barotropic magnetohydrodynamic equations:
\beq
 \vec \nabla \times (\hat {\vec v} \times \hat{\vec B}) = 0,
\label{Beqstat}
\enq
\beq
\rho (\hat {\vec v} \cdot \vec \nabla) \hat {\vec v} = -\vec \nabla p (\rho,s) +
\frac{(\vec \nabla \times \hat{\vec B}) \times \hat{\vec B}}{4 \pi}.
\label{Eulerstat}
\enq
In what follows we will attempt to reduce the number of variational variables from eight to four.

\section{Load and Metage}
\label{inverse}

\begin{figure}
\centering
\includegraphics[scale=0.5]{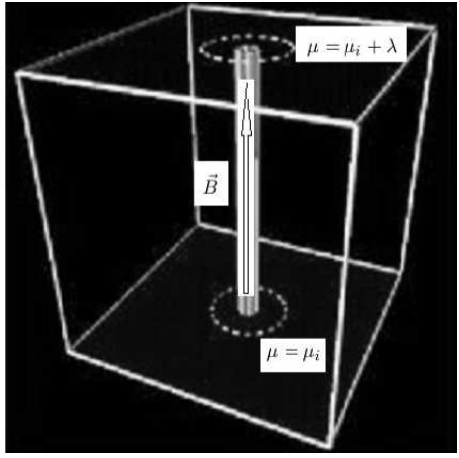}
\caption {A thin tube surrounding a magnetic field line}
\label{load}
\end{figure}

The following section follows closely a similar section in \cite{YaLy}. Consider a thin tube surrounding a magnetic field line as described in figure \ref{load},
the magnetic flux contained within the tube is:
\beq
\Delta \Phi =
\int \vec B \cdot d \vec S
\label{flux}
\enq
and the mass
contained with the tube is:
\beq
\Delta M = \int \rho d\vec l \cdot d \vec S,
\label{Mass}
\enq
in which $dl$ is a length element
along the tube. Since the magnetic field lines move with the flow
by virtue of \ern{Beq} and \ern{masscon} both the quantities $\Delta \Phi$ and
$\Delta M$ are conserved and since the tube is thin we may define
the conserved magnetic load:
\beq
\lambda = \frac{\Delta M}{\Delta
\Phi} = \oint \frac{\rho}{B}dl,
\label{Load}
\enq
in which the
above integral is performed along the field line. Obviously the
parts of the line which go out of the flow to regions in which
$\rho=0$ have a null contribution to the integral.
Notice that $\lambda$ is a {\bf single valued} function that can be measured in principle.
Since $\lambda$
is conserved it satisfies the equation:
\beq
 \frac{d \lambda }{d t} = 0.
\label{Loadcon}
\enq
By construction surfaces of constant magnetic load move with the flow and contain
magnetic field lines. Hence the gradient to such surfaces must be orthogonal to
the field line:
\beq
\vec \nabla \lambda \cdot \vec B = 0.
\label{Loadortho}
\enq
Now consider an arbitrary comoving point on the magnetic field line and denote it by $i$,
and consider an additional comoving point on the magnetic field line and denote it by $r$.
The integral:
\beq
\mu(r)  = \int_i^r \frac{\rho}{B}dl + \mu(i),
\label{metage}
\enq
is also a conserved quantity which we may denote following Lynden-Bell \& Katz \cite{LynanKatz}
as the magnetic metage. $\mu(i)$ is an arbitrary number which can be chosen differently for each
magnetic line. By construction:
\beq
 \frac{d \mu }{d t} = 0.
\label{metagecon}
\enq
Also it is easy to see that by differentiating along the magnetic field line we obtain:
\beq
 \vec \nabla \mu \cdot \vec B = \rho.
\label{metageeq}
\enq
Notice that $\mu$ will be generally a {\bf non single valued} function, we will show later in this paper
that symmetry to translations in $\mu$ will generate through the Noether theorem the conservation of the
magnetic cross helicity.

\begin{figure}
\centering
\includegraphics[scale=0.5]{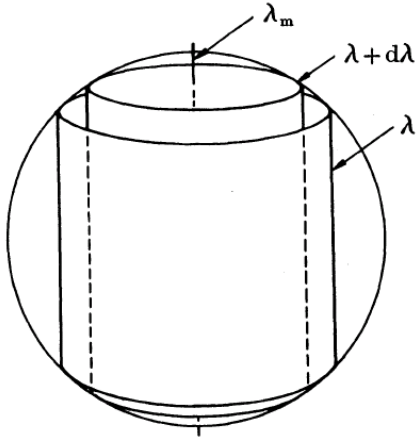}
\caption {Surfaces of constant load}
\label{loadsurface}
\end{figure}

At this point we have two comoving coordinates of flow, namely $\lambda,\mu$ obviously in a
three dimensional flow we also have a third coordinate. However, before defining the third coordinate
we will find it useful to work not directly with $\lambda$ but with a function of $\lambda$.
Now consider the magnetic flux within a surface of constant load $\Phi(\lambda)$ as described in figure \ref{loadsurface}
(the figure was given by Lynden-Bell \& Katz \cite{LynanKatz}). The magnetic flux is a conserved quantity
and depends only on the load $\lambda$ of the surrounding surface. Now we define the quantity:
\beq
 \chi = \frac{\Phi(\lambda)}{2\pi}.
\label{chidef}
\enq
Obviously $\chi$ satisfies the equations:
\beq
\frac{d \chi}{d t} = 0, \qquad \vec B \cdot \vec \nabla \chi = 0.
\label{chieq}
\enq
Let us now define an additional comoving coordinate $\eta^{*}$
since $\vec \nabla \mu$ is not orthogonal to the $\vec B$ lines we can choose $\vec \nabla \eta^{*}$ to be
orthogonal to the $\vec B$ lines and not be in the direction of the $\vec \nabla \chi$ lines,
that is we choose $\eta^{*}$
not to depend only on $\chi$. Since both $\vec \nabla \eta^{*}$ and $\vec \nabla \chi$ are orthogonal to $\vec B$,
$\vec B$ must take the form:
\beq
\vec B = A \vec \nabla \chi \times \vec \nabla \eta^{*}.
\enq
However, using \ern{Bcon} we have:
\beq
\vec \nabla \cdot \vec B = \vec \nabla A \cdot (\vec \nabla \chi \times \vec \nabla \eta^{*})=0.
\enq
Which implies that $A$ is a function of $\chi,\eta^{*}$. Now we can define a new comoving function
$\eta$ such that:
\beq
\eta = \int_0^{\eta^{*}}A(\chi,\eta^{'*})d\eta^{'*}, \qquad \frac{d \eta}{d t} = 0.
\enq
In terms of this function we obtain the Sakurai (Euler potentials) presentation:
\beq
\vec B = \vec \nabla \chi \times \vec \nabla \eta.
\label{Bsakurai3}
\enq
Hence we have shown how $\chi,\eta$ can be constructed for a known $\vec B,\rho$.
Notice however, that $\eta$ is defined in a non unique way since one can redefine
$\eta$ for example by performing the following transformation: $\eta \rightarrow \eta + f(\chi)$
in which $f(\chi)$ is an arbitrary function.
The comoving coordinates $\chi,\eta$ serve as labels of the magnetic field lines.
Moreover the magnetic flux can be calculated as:
\beq
\Phi = \int \vec B \cdot d \vec S = \int d \chi d \eta.
\label{phichieta}
\enq
In the case that the surface integral is performed inside a load contour we
obtain:
\beq
\Phi (\lambda) = \int_{\lambda} d \chi d \eta= \chi \int_{\lambda} d \eta =\left\{
\begin{array}{c}
 \chi [\eta] \\
  \chi (\eta_{max}-\eta_{min}) \\
\end{array}
\right.
\enq
 There are two cases involved; in one case the load surfaces are topological cylinders,
 in this case $\eta$ is not single valued and hence we obtain the upper value for $\Phi (\lambda)$.
 In a second case the load surfaces are topological spheres, in this case $\eta$ is single valued
 and has minimal $\eta_{min}$ and maximal $\eta_{max}$  values. Hence the lower value of $\Phi (\lambda)$ is obtained.
 For example in some cases $\eta$ is identical to twice the latitude angle $\theta$.
 In those cases $\eta_{min}=0$ (value at the "north pole") and $\eta_{max}= 2 \pi$
 (value at the "south pole").

Comparing the above \eq \ with \ern{chidef} we derive that $\eta$ can be either
{\bf single valued} or {\bf not single valued} and that
its discontinuity across its cut in the non single valued case is $[\eta] =2 \pi$.

So far the discussion did not differentiate the cases of stationary and non-stationary flows.
It should be noted that even for stationary flows one can have a non-stationary $\eta$ coordinates as the magnetic
field depends only on the gradient of $\eta$ (see \ern{Bsakurai3}), in particular if $\eta$ is stationary than  $\eta + g(t)$ which is clearly not stationary will produce according to \ern{Bsakurai3} a stationary magnetic field. In what follows we find it advantageous
to use the form of $\eta$ given in \ern{etastas} in which $\bar \eta$ is stationary.

The triplet $\chi,\eta,\mu$ will suffice to label any fluid element in three dimensions. But for a non-barotropic
flow there is also another possible label $s$ which is comoving according to \ern{Ent}. The question then arises of the relation of this
label to the previous three. As one needs to make a choice regarding the preferred set of labels it seems that the physical
ones are $\chi,\eta,s$ in which we use the surfaces on which the magnetic fields lie and the entropy, each label has an obvious
physical interpretation. In this case we must look at $\mu$ as a function of $\chi,\eta,s$. If the magnetic field lines lie on entropy
surface then $\mu$ regains its status as an independent label. The density can now be written as:
\beq
\rho = \frac{\partial \mu}{ \partial s} \frac{\partial(\chi,\eta,s)}{\partial(x,y,z)}.
\label{metagecon3}
\enq
Now as $\mu$ can be defined for each magnetic field line separately according to \ern{metage} it is obvious that
such a choice exist in which $\mu$ is a function of $s$ only. One may also think of the entropy $s$ as a functions
$\chi,\eta,\mu$. However, if one change $\mu$ in this case this generally entails a change in $s$ and the symmetry described in
\ern{metage} is lost in the Action. In what follows we shall ignore the status of $s$ as a label and consider it as a variational variable which only attains a status of a label at the variational extremum.

\section{A Simpler variational principle of stationary \\ non-barotropic magnetohydrodynamics}

In a previous paper \cite{Yahalomc} we have shown that stationary non-barotropic magnetohydrodynamics
can be described in terms of eight first order differential equations and
 by an action principle from which those equations can be derived.
Below we will show that one can do better for the case in which the magnetic field lines
lie on an entropy surface, in this case three functions will suffice to describe stationary non-barotropic magnetohydrodynamics.

Consider \ern{chieq}, for a stationary flow it takes the form:
\beq
\vec v \cdot \vec \nabla \chi = 0.
\label{chieqstat}
\enq
Hence $\vec v$ can take the form:
\beq
\vec v = \frac{\vec \nabla \chi \times \vec K}{\rho}.
\label{orthov}
\enq
However, the velocity field must satisfy the stationary mass conservation \ern{masscon}:
\beq
\vec \nabla \cdot (\rho \vec v ) = 0.
\label{massconstat}
\enq
We see that a sufficient condition (although not necessary)
for $\vec v$ to solve \ern{massconstat} is that $\vec K$ takes the
form $\vec K = \vec \nabla N$, where $N$ is an arbitrary function. Thus,
$\vec v$ may take the form:
\beq
\vec v = \frac{\vec \nabla \chi \times \vec \nabla N}{\rho}.
\label{orthov2}
\enq
Let us now calculate $\vec v \times \vec B$ in which $\vec B$ is given by Sakurai's presentation
\ern{Bsakurai3}:
\ber
\vec v \times \vec B &=& (\frac{\vec \nabla \chi \times \vec \nabla N}{\rho}) \times
(\vec \nabla \chi \times \vec \nabla \eta)
\nonumber \\
&=& \frac{1}{\rho} \vec \nabla \chi (\vec \nabla \chi \times \vec \nabla N) \cdot \vec \nabla \eta.
\label{orthov3}
\enr
Since the flow is stationary $N$ can be at most a function of the three comoving coordinates
$\chi,\mu,\bar \eta$ defined in section \ref{inverse}, hence:
\beq
\vec \nabla N  = \frac{\partial N}{\partial \chi} \vec \nabla \chi +
\frac{\partial N}{\partial \mu} \vec \nabla \mu + \frac{\partial N}{\partial \bar \eta} \vec \nabla \bar \eta.
\label{Ndiv}
\enq
Inserting \ern{Ndiv} into \er{orthov3} will yield:
\beq
\vec v \times \vec B =
\frac{1}{\rho} \vec \nabla \chi \frac{\partial N}{\partial \mu}
(\vec \nabla \chi \times \vec \nabla \mu) \cdot \vec \nabla \bar \eta.
\label{orthov4}
\enq
Rearranging terms and using Sakurai's presentation \ern{Bsakurai3} we can
simplify the above equation and obtain:
\beq
\vec v \times \vec B = -\frac{1}{\rho} \vec \nabla \chi \frac{\partial N}{\partial \mu}
(\vec \nabla \mu \cdot \vec B).
\label{orthov5}
\enq
However, using \ern{metageeq} this will simplify to the form:
\beq
\vec v \times \vec B = - \vec \nabla \chi \frac{\partial N}{\partial \mu}.
\label{orthov51}
\enq
Inserting \ern{orthov51} into \ern{Beqs} will lead to the equation:
\beq
\vec \nabla  (\frac{\partial N}{\partial \mu}) \times \vec \nabla \chi = 0.
\label{Beqstatimp}
\enq
However, since $N$ is at most a function of $\chi,\mu,\bar \eta$ it follows that
$\frac{\partial N}{\partial \mu}$ is some function of $\chi$:
\beq
\frac{\partial N}{\partial \mu} = -F(\chi).
\enq
This can be easily integrated to yield:
\beq
N = - \mu F(\chi) + G(\chi,\bar \eta).
\enq
Inserting this back into \ern{orthov2} will yield:
\beq
\vec v = \frac{\vec \nabla \chi \times (-F(\chi) \vec \nabla \mu+
\frac{\partial G}{\partial \bar \eta} \vec \nabla \bar \eta) }{\rho}.
\label{orthov7}
\enq
Let us now replace the set of variables $\chi,\bar \eta$ with a new set $\chi',\bar \eta'$
such that:
\beq
\chi' = \int F(\chi) d\chi, \qquad \bar \eta' = \frac{\bar \eta}{F(\chi)}.
\enq
This will not have any effect on the Sakurai representation given in \ern{Bsakurai3} since:
\beq
\vec B = \vec \nabla \chi \times \vec \nabla \eta = \vec \nabla \chi \times \vec \nabla \bar \eta =
\vec \nabla \chi' \times \vec \nabla \bar \eta'.
\enq
However, the velocity will have a simpler representation and will take the form:
\beq
\vec v = \frac{\vec \nabla \chi' \times \vec \nabla(- \mu+ G'(\chi',\bar \eta'))}{\rho},
\label{orthov8}
\enq
in which $G'=\frac{G}{F}$. At this point one should remember that $\mu$ was defined in \ern{metage}
up to an arbitrary constant which can vary between magnetic field lines. Since the lines
are labelled by their $\chi',\bar \eta'$ values it follows that we can add an arbitrary function of
$\chi',\bar \eta'$ to $\mu$ without effecting its properties. Hence we can define a new $\mu'$ such that:
\beq
\mu' =\mu-  G'(\chi',\bar \eta').
\label{mupdef}
\enq
Notice that $\mu'$ can be multi-valued. Inserting \er{mupdef} into \er{orthov8} will lead to a simplified equation for $\vec v$:
\beq
\vec v = \frac{\vec \nabla \mu' \times \vec \nabla \chi'}{\rho}.
\label{orthov9}
\enq
In the following the primes on $\chi,\mu,\bar \eta$ will be ignored.
The above equation is analogues to Vladimirov and Moffatt's \cite{Moffatt} equation 7.11
for incompressible flows, in which our $\mu$ and $\chi$ play the part of their $A$ and
$\Psi$. It is obvious that
$\vec v$ satisfies the following set of equations:
\beq
\vec v \cdot \vec \nabla \mu = 0, \qquad  \vec v \cdot \vec \nabla \chi = 0, \qquad
\vec v \cdot \vec \nabla \bar \eta = 1,
\label{comov}
\enq
to derive the right hand equation we have used both \ern{metagecon} and \ern{Bsakurai3}.
Hence $\mu,\chi$ are both comoving and stationary. As for $\bar \eta$ it satisfies the same
equation as $\bar \eta$ defined in \ern{etastas}.
It can be easily seen that if:
\beq
basis = (\vec \nabla \chi, \vec \nabla \bar \eta,\vec \nabla \mu),
\enq
is a local vector basis at any point in space than their exists a dual basis:
\beq
dual \ basis = \frac{1}{\rho}(\vec \nabla \bar \eta \times \vec \nabla \mu, \vec \nabla \mu \times \vec \nabla \chi
, \vec \nabla \chi \times \vec \nabla \bar \eta ) =
(\frac{\vec \nabla \bar \eta \times \vec \nabla \mu}{\rho}, \vec v, \frac{\vec B}{\rho} ).
\enq
Such that:
\beq
basis_i \cdot dual \ basis_j = \delta_{ij}, \qquad i,j\in[1,2,3],
\enq
in which $\delta_{ij}$ is Kronecker's delta.
Hence while the surfaces $\chi,\mu,\bar \eta$ generate a local vector basis for space, the
physical fields of interest $\vec v,\vec B$ are part of the dual basis.
By vector multiplying $\vec v$ and $\vec B$ and using equations (\ref{orthov9},\ref{Bsakurai3})
we obtain:
\beq
\vec v \times \vec B = \vec \nabla \chi,
\label{vBsurfaces}
\enq
this means that both $\vec v$ and $\vec B$ lie on $\chi$ surfaces and provide a vector
basis for this two dimensional surface. The above equation can be compared
with Vladimirov and Moffatt \cite{Moffatt} equation 5.6 for incompressible flows in which their
$J$ is analogue to our $\chi$.

\section{The three function action principle for stationary flows}

In the previous subsection we have shown that if the velocity field $\vec v$
is given by \ern{orthov9} and the magnetic field $\vec B$ is given by the Sakurai
representation \ern{Bsakurai3} than \eqs (\ref{Beqs},\ref{Bcons},\ref{masscons})
are satisfied automatically for stationary flows. To complete the set of equations
we will show how the Euler \ers{Euler} can be derived from the action:
\ber
A & \equiv & \int {\cal L} d^3 x dt,
\nonumber \\
{\cal L} & \equiv & \rho (\frac{1}{2} \vec v^2 - \varepsilon (\rho,s)) - \frac{\vec B^2}{8 \pi},
\label{Lagaction}
\enr
in which both $\vec v$ and $\vec B$ are given by \ern{orthov9} and \ern{Bsakurai3} respectively
and the density $\rho$ is given by \ern{metagecon}:
\beq
\rho = \vec \nabla \mu \cdot \vec B = \vec \nabla \mu \cdot (\vec \nabla \chi \times \vec \nabla \eta)
=\frac{\partial(\chi,\eta,\mu)}{\partial(x,y,z)}.
\label{metagecon2}
\enq

The Lagrangian density of \ern{Lagaction} takes the more explicit form:
\beq
{\cal L}[\chi,\eta,\mu] = \rho \left(\frac{1}{2} (\frac{\vec \nabla \mu \times \vec \nabla \chi}{\rho})^2 - \varepsilon (\rho,s(\chi,\eta,\mu))\right) - \frac{(\vec \nabla \chi \times \vec \nabla \eta)^2}{8 \pi}
\label{lagstatsimp}
\enq
and can be seen explicitly to depend on only three functions. We underline that if the magnetic
field lines lie on entropy surfaces. $s$ must be a function of $\chi,\eta$ only and does not depend on $\mu$.
Let us make arbitrary small variations $\delta \alpha_i =(\delta \chi,\delta \eta,\delta \mu)$ of the functions
$\alpha_i=(\chi,\eta,\mu)$. Let us define a $\Delta$ variation that does not modify the $\alpha_i$'s, such that:
\beq
\Delta \alpha_i = \delta \alpha_i + (\vec \xi \cdot \vec \nabla) \alpha_i = 0,
\label{Delal}
\enq
in which $\vec \xi$ is the Lagrangian displacement, thus:
\beq
\delta \alpha_i = -\vec \nabla \alpha_i \cdot \vec \xi.
\label{delal}
\enq
Which will lead to the equation:
\beq
\vec \xi \equiv -\frac{\partial \vec r}{\partial \alpha_i} \delta \alpha_i.
\label{xialdef}
\enq
 Making a variation of $\rho$ given in \ern{metagecon2} with respect to $\alpha_i$ will yield:
 \beq
\delta \rho = - \vec \nabla \cdot (\rho \vec \xi).
\label{delrho}
\enq
Making a variation of $s$ will result in:
\beq
\delta s = \frac{\partial s}{\partial \alpha_i} \delta \alpha_i = - \frac{\partial s}{\partial \alpha_i}  \vec \nabla \alpha_i \cdot \vec \xi  = -  \vec \nabla s \cdot \vec \xi.
\label{dels}
\enq

 Furthermore, taking the variation of $\vec B$ given by Sakurai's representation
(\ref{Bsakurai3}) with respect to $\alpha_i$ will yield:
\beq
\delta \vec B = \vec \nabla \times (\vec \xi \times \vec B).
\label{delB}
\enq
It remains to calculate $\delta \vec v$ by varying \ern{orthov9} this will yield:
\beq
\delta \vec v = -\frac{\delta \rho}{\rho} \vec v + \frac{1}{\rho} \vec \nabla \times (\rho \vec \xi \times \vec v).
\label{delv2}
\enq
Varying the action will result in:
\ber
\delta A & = & \int \delta {\cal L} d^3 x dt,
\nonumber \\
\delta {\cal L} & = &  \delta \rho (\frac{1}{2} \vec v^2 - w (\rho,s))- \rho T \delta s
+\rho \vec v \cdot \delta  \vec v - \frac{\vec B \cdot \delta \vec B}{4 \pi},
\label{delLagaction}
\enr
Inserting \eqs (\ref{delrho},\ref{delB},\ref{delv2}) into \ern{delLagaction}
will yield:
\ber
\delta {\cal L} &=& \vec v \cdot \vec \nabla \times (\rho \vec \xi \times \vec v)
- \frac{\vec B \cdot \vec \nabla \times (\vec \xi \times \vec B)}{4 \pi}
- \delta \rho (\frac{1}{2} \vec v^2 + w) + \rho T \vec \nabla s \cdot \vec \xi
\nonumber \\
&=& \vec v \cdot \vec \nabla \times (\rho \vec \xi \times \vec v)
- \frac{\vec B \cdot \vec \nabla \times (\vec \xi \times \vec B)}{4 \pi}
+ \vec \nabla \cdot (\rho \vec \xi) (\frac{1}{2} \vec v^2 + w )
\nonumber \\
&+& \rho T \vec \nabla s \cdot \vec \xi.
\label{delcalL}
\enr
Using the well known vector identity:
\beq
\vec A \cdot \vec \nabla \times (\vec C \times \vec A)=
\vec \nabla \cdot ((\vec C \times \vec A) \times \vec A)+
(\vec C \times \vec A) \cdot \vec \nabla \times \vec A
\label{veciden1}
\enq
and the theorem of Gauss we can write now \ern{delLagaction} in the form:
\ber
\delta A & = &
\int dt \{ \oint d \vec S \cdot [\rho (\vec \xi \times \vec v)\times \vec v
-\frac{(\vec \xi \times \vec B)\times \vec B}{4 \pi}+(\frac{1}{2} \vec v^2 + w )\rho \vec \xi]
\nonumber \\
& + & \int d^3 x  \vec \xi \cdot [\rho \vec v \times \vec \omega+
\vec J \times \vec B-\rho \vec \nabla (\frac{1}{2} \vec v^2 + w ) + \rho T \vec \nabla s]\}.
\label{delLagaction2simpl}
\enr
The time integration is of course redundant in the above expression. Also notice that we have used
the current definition \ern{J} and the vorticity definition $ \vec \omega = \vec \nabla \times \vec v$. Suppose now
that $\delta A = 0$ for a $\vec \xi$ such that the boundary term in the above equation
is null but that $\vec \xi$ is otherwise arbitrary, then it entails the equation:
\beq
\rho \vec v \times \vec \omega+
\vec J \times \vec B-\rho \vec \nabla (\frac{1}{2} \vec v^2 + w ) + \rho T \vec \nabla s = 0.
\enq
Using the well known vector identity:
\beq
\frac{1}{2} \vec \nabla (\vec v^2) = (\vec v \cdot \vec \nabla) \vec v +
 \vec v \times (\vec \nabla \times \vec v)
\label{veciden2}
\enq
and rearranging terms we recover the stationary Euler equation:
\beq
\rho (\vec v \cdot \vec \nabla)\vec v = -\vec \nabla p +  \vec J \times \vec B.
\label{Eulerstat2}
\enq

\section{The three function action principle for a static configuration}

The static configuration is a stationary flow such that $\vec v =0$ . In this case
the mass conservation \ern{Beqs} and magnetic field \ern{masscons} are
satisfied trivially.   To complete the set of equations
we will show how the static Euler \ers{Euler} can be derived from the action:
\ber
A & \equiv & -\int {\cal L} d^3 x dt,
\nonumber \\
{\cal L} & \equiv & \rho \varepsilon (\rho,s) + \frac{\vec B^2}{8 \pi},
\label{Lagactionstat}
\enr
in which  $\vec B$ is given by \ern{Bsakurai3} and the density $\rho$ is given by \ern{metagecon2}.
The Lagrangian density of \ern{Lagactionstat} can be put in the more explicit form:
\beq
{\cal L}[\chi,\eta,\mu] = \rho \varepsilon (\rho,s(\chi,\eta,\mu))+\frac{(\vec \nabla \chi \times \vec \nabla \eta)^2}{8 \pi}
\label{lagstatsimpstat}
\enq
Varying the action will result in:
\ber
\delta A & = & -\int \delta {\cal L} d^3 x dt,
\nonumber \\
\delta {\cal L} & = &  \delta \rho (w (\rho,s))+ \rho T \delta s
+ \frac{\vec B \cdot \delta \vec B}{4 \pi},
\label{delLagactionstat}
\enr
Inserting \eqs (\ref{delrho},\ref{delB}) into \ern{delLagactionstat}
will yield:
\ber
\delta {\cal L} &=&
 \frac{\vec B \cdot \vec \nabla \times (\vec \xi \times \vec B)}{4 \pi}
+ \delta \rho  w + \rho T \vec \nabla s \cdot \vec \xi
\nonumber \\
&=& \frac{\vec B \cdot \vec \nabla \times (\vec \xi \times \vec B)}{4 \pi}
- \vec \nabla \cdot (\rho \vec \xi)  w
+ \rho T \vec \nabla s \cdot \vec \xi.
\label{delcalLst}
\enr
Using the well known vector identity (\ref{veciden1}),
and the theorem of Gauss we can write now \ern{delLagactionstat} in the form:
\ber
\delta A & = &
\int dt \{ \oint d \vec S \cdot [
-\frac{(\vec \xi \times \vec B)\times \vec B}{4 \pi}+ w\rho \vec \xi]
\nonumber \\
& + & \int d^3 x  \vec \xi \cdot [
\vec J \times \vec B-\rho \vec \nabla w + \rho T \vec \nabla s]\}.
\label{delLagaction2simplstat}
\enr
The time integration is of course redundant in the above expression. Also notice that we have used
the current definition \ern{J}. Suppose now
that $\delta A = 0$ for a $\vec \xi$ such that the boundary term in the above equation
is null but that $\vec \xi$ is otherwise arbitrary, then it entails the equation:
\beq
\vec J \times \vec B-\rho \vec \nabla w  + \rho T \vec \nabla s = 0.
\enq
and rearranging terms we recover the stationary Euler equation:
\beq
 \vec J \times \vec B - \vec \nabla p = 0.
\label{Eulerstat2stat}
\enq

\section{Transport phenomena}

In many plasmas including static configurations heat is transferred preferably along magnetic field
lines:
\beq
 \vec J_H = - \hat k \vec \nabla T .
\label{Heatrans}
\enq
in which $\hat k$ is a tensor of heat conductivity. This tensor is usually larger in the magnetic field direction and thus can be written as:
\beq
  \hat k  = k_\bot(I- \hat b \otimes \hat b) + k_\| \hat b \otimes \hat b .
\label{Heatrans2}
\enq
in the above $\hat b \equiv \frac{\vec B}{B}$ is a unit vector in the magnetic field direction, $\otimes$ is the tensor product, $I$ is the unit matrix, $k_\bot$ is the heat conductivity in directions perpendicular to magnetic field lines
and  $k_\|$ is the larger heat conductivity in the direction parallel to magnetic field lines.
The equation for a stationary heat flux configuration is:
\beq
 \vec \nabla \cdot \vec J_H = 0 \Rightarrow  \vec \nabla \cdot (\hat k \vec \nabla T) = 0.
\label{Heatranseq}
\enq
This equation can be derived from the heat Lagrangian \&  Lagrangian density:
\beq
L_H \equiv \int {\cal L}_H d^3 x, \qquad  {\cal L}_H \equiv \frac{1}{2} (\vec \nabla T)^t \hat k \vec \nabla T =
  \frac{1}{2} \partial_i T \hat k_{ij} \partial_j T,
\label{Heatlagden}
\enq
in the above $(\vec \nabla T)^t$ is the transpose of the $\vec \nabla T$, $\partial_i \equiv \frac{\partial}{\partial x_i}$ and Einstein summation convention is assumed. Taking the variation
with respect to the temperature $T$ yields:
\beq
\delta {\cal L}_H =  (\vec \nabla T)^t \hat k \vec \nabla \delta T ,
\label{delHeatlagden}
\enq
hence:
\beq
\delta L_H = \int d^3 x \left[\vec \nabla \cdot (\hat k \vec \nabla T \delta T)
 - \delta T \vec \nabla \cdot (\hat k \vec \nabla T ) \right],
\label{delHeatlagden2}
\enq
and using the theorem of Gauss:
\beq
\delta L_H = \int d \vec S \cdot (\hat k \vec \nabla T) \delta T
 - \int d^3 x \delta T \vec \nabla \cdot (\hat k \vec \nabla T ) ,
\label{delHeatlagden3}
\enq
thus for appropriate boundary conditions we derive \ern{Heatranseq}. We notice that heat
conduction is not taken into account in ideal MHD which only assumes convection of heat.
However, provided that conduction is seen as a secondary process with respect to convection
we may obtain using the ideal variational principle a stationary or static magnetic field configuration
using the appropriate variational expression given in previous sections. And then using the known magnetic field  configuration we derive the appropriate heat flux transport using $L_H$.

\section {Conclusion}

It is shown that stationary non-barotropic magnetohydrodynamics can be derived from a variational principle of three functions. We have shown this for both the stationary and static cases.

Possible applications include stability analysis of stationary magnetohydrodynamic configurations and its possible utilization
for developing efficient numerical schemes for integrating the magnetohydrodynamic equations.
It may be more efficient to incorporate the developed formalism in the frame work of an existing
code instead of developing a new code from scratch. Possible existing codes are described
in \cite{Mignone,Miyoshi,Narayan,Shapiro,Reisenegger}.
We anticipate applications of this study both to linear and non-linear stability analysis of known
 barotropic magnetohydrodynamic configurations \cite{VMI,Kruskal,AHH}.
 We suspect that for achieving this we will need to add additional
constants of motion constraints to the action as was done by \cite{Arnold1,Arnold2}
see also \cite{Katz,YahalomKatz,YahalomMonth}. As for designing efficient numerical schemes for integrating
the equations of fluid dynamics and magnetohydrodynamics one may follow the
approach described in \cite{Yahalom,YahalomPinhasi,YahPinhasKop,OphirYahPinhasKop}.

Another possible application of the variational method is in deducing new analytic
solutions for the magnetohydrodynamic equations. Although the equations are notoriously
difficult to solve being both partial differential equations and nonlinear, possible
solutions can be found in terms of variational variables. An example
for this approach is the self gravitating torus described in \cite{Yah3}.

One can use continuous symmetries which appear in the variational Lagrangian to derive
through Noether theorem new conservation laws. An example for such derivation which
still lacks physical interpretation can be found in \cite{Yah5}. It may be that the
Lagrangian derived in \cite{Yah} has a larger symmetry group. And of course one anticipates
a different symmetry structure for the non-barotropic case.

Topological invariants have always been informative, and there are such invariants in MHD flows. For
example the two helicities  have long been useful in research into
the problem of hydrogen fusion, and in various astrophysical scenarios. In previous
works \cite{YaLy,Yah2,Yahalomhel} connections between helicities with symmetries of the barotropic fluid equations were made. The variables of
the current variational principles are helpful for identifying and characterizing new topological invariants in MHD \cite{Yahalomhel2,Yahalomhel3,Yahalomhel4}.

Although ideal MHD does not describe fully real plasmas, we show here how processes such as heat conduction  can be also described using variational analysis provided that the magnetic field configuration is given approximately by ideal variational analysis.


\begin{thebibliography}{}

\bibitem {Sturrock}
P. A.  Sturrock, {\it Plasma Physics} (Cambridge University Press,
Cambridge, 1994)
\bibitem{Moffatt}
V. A. Vladimirov and H. K. Moffatt, J. Fluid. Mech. {\bf 283}
125-139 (1995)
\bibitem {Kats}
A. V. Kats, Los Alamos Archives physics-0212023 (2002), JETP Lett.
77, 657 (2003)
\bibitem {Kats3}
A. V. Kats and V. M. Kontorovich, Low Temp. Phys. 23, 89 (1997)
\bibitem {Kats4}
A. V. Kats, Physica D 152-153, 459 (2001)
\bibitem {Sakurai}
T. Sakurai,  Pub. Ast. Soc. Japan {\bf 31} 209 (1979)
\bibitem{Morrison}
P.J. Morrison, Poisson Brackets for Fluids and Plasmas, AIP Conference proceedings, Vol. 88, Table 2.
\bibitem {Yang}
W. H. Yang, P. A. Sturrock and S. Antiochos, Ap. J., {\bf 309} 383 (1986)
\bibitem {YaLy}
 A. Yahalom and D. Lynden-Bell, "Simplified Variat\-ional Princi\-ples for Barotropic Magnetohydrodynamics,"\\ (Los-Alamos Archives \-
physics/0603128) {\it Journal of Fluid Mechanics}, Vol. 607, 235-265, 2008.
\bibitem {Yah}
Yahalom A., "A Four Function Variational Principle for Barotropic Magnetohydrodynamics"
EPL 89 (2010) 34005, doi: 10.1209/0295-5075/89/34005 [Los - Alamos Archives - arXiv: 0811.2309]
\bibitem {Yah2}
Asher Yahalom "Aharonov - Bohm Effects in Magnetohydrodynamics" Physics Letters A.
Volume 377, Issues 31-33, 30 October 2013, Pages 1898-1904.
\bibitem {FLS}
A. Frenkel, E. Levich and L. Stilman Phys. Lett. A {\bf 88}, p.
461 (1982)
\bibitem {Zakharov}
V. E. Zakharov and E. A. Kuznetsov, Usp. Fiz. Nauk 40, 1087 (1997)
\bibitem{Bekenstien}
J. D. Bekenstein and A. Oron, Physical Review E Volume 62, Number
4, 5594-5602 (2000)
\bibitem {Kats2}
A. V. Kats, Phys. Rev E 69, 046303 (2004)
\bibitem {Yahalomnb}
A. Yahalom "Simplified Variational Principles for non-Barotropic Magnetohydrodynamics". (arXiv: 1510.00637 [Plasma Physics]) J. Plasma Phys. (2016), vol. 82, 905820204. doi:10.1017/S0022377816000222.
\bibitem {Yahalomnb2}
Asher Yahalom "Non-Barotropic Magnetohydrodynamics as a Five Function Field Theory". International Journal of Geometric Methods in Modern Physics, No. 10 (November 2016). Vol. 13 1650130 (13 pages) \copyright World Scientific Publishing Company.
\bibitem {Yahalomc}
A. Yahalom "Simplified Variational Principles for Stationary non-Barotropic Magnetohydrodynamics" International Journal of Mechanics, Volume 10, 2016, p. 336-341. ISSN: 1998-4448.
\bibitem {Yahalomd}
Asher Yahalom "A Simpler Variational Principle for Stationary non-Barotropic Ideal Magnetohydrodynamics". Chaotic Modeling and Simulation (CMSIM), 1: 19-33, 2018. Received: 15 October 2017 / Accepted: 28 December 2017.
\bibitem{LynanKatz}
D. Lynden-Bell and J. Katz "Isocirculational Flows and their Lagrangian and Energy principles",
Proceedings of the Royal Society of London. Series A, Mathematical and Physical Sciences, Vol. 378,
No. 1773, 179-205 (Oct. 8, 1981).
\bibitem {Mignone}
Mignone, A., Rossi, P., Bodo, G., Ferrari, A., \& Massaglia, S. (2010). High-resolution 3D relativistic MHD simulations of jets.
 Monthly Notices of the Royal Astronomical Society, 402(1), 7-12.
 \bibitem {Miyoshi}
 Miyoshi, T., \& Kusano, K. (2005). A multi-state HLL approximate Riemann solver for ideal
 magnetohydrodynamics. Journal of Computational Physics, 208(1), 315-344.
\bibitem {Narayan}
 Igumenshchev, I. V., Narayan, R., \& Abramowicz, M. A. (2003). Three-dimensional magnetohydrodynamic simulations of radiatively
  inefficient accretion flows. The Astrophysical Journal, 592(2), 1042.
\bibitem {Shapiro}
 Faber, J. A., Baumgarte, T. W., Shapiro, S. L., \& Taniguchi, K. (2006). General relativistic binary merger simulations and
 short gamma-ray bursts. The Astrophysical Journal Letters, 641(2), L93.
\bibitem {Reisenegger}
 Hoyos, J., Reisenegger, A., \& Valdivia, J. A. (2007). Simulation of the Magnetic Field Evolution in Neutron Stars.
  In VI Reunion Anual Sociedad Chilena de Astronomia (SOCHIAS) (Vol. 1, p. 20).
\bibitem{VMI}
V. A. Vladimirov, H. K. Moffatt and K. I. Ilin, J. Fluid Mech.
329, 187 (1996); J. Plasma Phys. 57, 89 (1997); J. Fluid Mech. 390, 127 (1999)
\bibitem {Kruskal}
 Bernstein, I. B., Frieman, E. A., Kruskal, M. D., \& Kulsrud, R. M. (1958).
 An energy principle for hydromagnetic stability problems. Proceedings of the Royal
 Society of London. Series A. Mathematical and Physical Sciences, 244(1236), 17-40.
\bibitem{AHH}
J. A. Almaguer, E. Hameiri, J. Herrera, D. D. Holm, Phys. Fluids,
31, 1930 (1988)
\bibitem{Arnold1}
V. I. Arnold, Appl. Math. Mech. {\bf 29}, 5, 154-163.
\bibitem{Arnold2}
V. I. Arnold, Dokl. Acad. Nauk SSSR {\bf 162} no. 5.
\bibitem{Katz}
J. Katz, S. Inagaki, and A. Yahalom,  "Energy Principles for
Self-Gravitating Barotropic Flows: I. General Theory", Pub. Astro. Soc. Japan 45, 421-430 (1993).
\bibitem{YahalomKatz}
Yahalom A., Katz J. \& Inagaki K. 1994, {\it Mon. Not. R. Astron. Soc.} {\bf 268} 506-516.
\bibitem{YahalomMonth}
A. Yahalom, "Stability in the Weak Variational Principle of Barotropic Flows and Implications for Self-Gravitating Discs".
 Monthly Notices of the Royal Astronomical Society 418, 401-426 (2011).
\bibitem{Yahalom}
A. Yahalom, "Method and System for Numerical Simulation of Fluid
Flow", US patent 6,516,292 (2003).
\bibitem{YahalomPinhasi}
A. Yahalom, \& G. A.  Pinhasi, "Simulating Fluid Dynamics using a
Variational Principle", proceedings of the AIAA Conference, Reno, USA (2003).
\bibitem{YahPinhasKop}
A. Yahalom, G. A. Pinhasi and M. Kopylenko, "A Numerical Model
Based on Variational Principle for Airfoil and Wing Aerodynamics",
proceedings of the AIAA Conference, Reno, USA (2005).
\bibitem{OphirYahPinhasKop}
D. Ophir, A. Yahalom, G.A. Pinhasi and M. Kopylenko "A Combined Variational and Multi-Grid Approach for Fluid Dynamics Simulation" Proceedings
of the ICE - Engineering and Computational Mechanics, Volume 165, Issue 1, 01 March 2012, pages 3 -14 , ISSN: 1755-0777, E-ISSN: 1755-0785.
\bibitem {Yah3}
Asher Yahalom "Using fluid variational variables to obtain new analytic
 solutions of self-gravitating flows with nonzero helicity" Procedia IUTAM 7 (2013) 223 - 232.
\bibitem {Bateman}
 H. Bateman "On Dissipative Systems and Related Variational Principles" Phys. Rev. 38, 815  Published 15 August 1931.
 \bibitem {Yah5}
Asher Yahalom, "A New Diffeomorphism Symmetry Group of Magnetohydrodynamics" V. Dobrev (ed.), Lie Theory and Its Applications in Physics:
IX International Workshop, Springer Proceedings in Mathematics \& Statistics 36, p. 461-468, 2013.
\bibitem{KLB}
Katz, J. \& Lynden-Bell, D. Geophysical \& Astrophysical Fluid
Dynamics 33,1 (1985).
\bibitem{bt}
Binney J. \& Tremaine S., 1987, Galactic Dynamics, Princeton University Press
\bibitem{Clebsch1}
Clebsch, A., Uber eine allgemeine Transform\-ation der hydro-\-dynamischen Gleichungen.
{\itshape J.~reine angew.~Math.}~1857, {\bf 54}, 293--312.
\bibitem{Clebsch2}
Clebsch, A., Uber die Integration der hydrodynamischen Gleichungen.
{\itshape J.~reine angew.~Math.}~1859, {\bf 56}, 1--10.
\bibitem{Yahalomhel}
A. Yahalom, "Helicity Conservation via the Noether Theorem" J.
Math. Phys. 36, 1324-1327 (1995). [Los-Alamos Archives solv-int/9407001]
\bibitem{Yahalomhel2}
Asher Yahalom "A Conserved Local Cross Helicity for Non-Barotropic MHD" (ArXiv 1605.02537). Pages 1-7, Journal of Geophysical\- \& Astrophys\-ical Fluid Dynamics. Published online: 25 Jan 2017. Vol. 111, No. 2, 131-137.
\bibitem{Yahalomhel3}
Asher Yahalom "Non-Barotropic Cross-helicity Conservation Applications in Magnetohydrodynamics and the Aharanov - Bohm effect" (arXiv:1703.08072 [physics.plasm-ph]). Fluid Dynamics Research, Volume 50, Number 1, 011406.  https://doi.org/10.1088/1873-7005/aa6fc7 . Received 11 December 2016, Accepted Manuscript online 27 April 2017, Published 30 November 2017.
\bibitem{Yahalomhel4}
Asher Yahalom \& Hong Qin "Noether Currents for Eulerian Variational Principles in Non-Barotropic Magnetohydrodynamics and Topological Conservations Laws" Journal of Fluid Mechanics, 908, A4. doi:10.1017/jfm.2020.856, 2021.

\end{thebibliography}
\end{document}